\documentclass[aps,prd,twocolumn,superscriptaddress,showkeys]{revtex4}
\usepackage{amsmath}
\usepackage{color,xcolor}
\usepackage{graphicx}

\newcommand{\bastar}{\begin{eqnarray*}}
\newcommand{\eastar}{\end{eqnarray*}}
\newskip\humongous \humongous=0pt plus 1000pt minus 1000pt

\newif\ifdtup

\relax
\newcommand{\W}{{\vec W}}
\newcommand{\n}{\hat n}
\newcommand{\hn}{\hat n}
\newcommand{\hr}{\hat r}

\newcommand{\hD}{{\hat D}}

\newcommand{\pd}{\partial}

\newcommand{\A}{{\vec A}}
\newcommand{\hA}{{\hat A}}

\newcommand{\tA}{{\tilde A}}
\newcommand{\tC}{{\tilde C}}
\newcommand{\F}{{\vec F}}

\newcommand{\hF}{{\hat F}}

\newcommand{\bg}{\bar g}
\newcommand{\mn}{{\mu\nu}}

\newcommand{\vsig}{{\vec \sigma}}
\newcommand{\nn}{\nonumber}

\newcommand{\cL}{{\cal L}}

\newcommand{\eps}{{\epsilon}}
\newcommand{\beps}{{\bar \epsilon}}
\newcommand{\teps}{\tilde \epsilon}
\begin{document}
\title{Regularization of Electroweak Monopole by Charge 
Screening and BPS Energy Bound}

\author{Pengming Zhang}
\email{zhangpm5@mail.sysu.edu.cn}
\affiliation{School of Physics and Astronomy,
Sun Yat-Sen University, Zhuhai 519082, China}
\author{Li-Ping Zou}
\email{zoulp@impcas.ac.cn}
\affiliation{Institute of Modern Physics, Chinese Academy 
of Science, Lanzhou 730000, China}
\author{Y. M. Cho}
\email{ymcho0416@gmail.com}
\affiliation{School of Physics and Astronomy,
Seoul National University, Seoul 08826, Korea}
\affiliation{Center for Quantum Spacetime, 
Sogang University, Seoul 04107, Korea} 

\begin{abstract}
We show that the electroweak monopole can be 
regularized with a non-vacuum electromagnetic 
permittivity. This allows us to set a new BPS bound 
for the monopole mass, which implies that the mass 
may not be smaller than 2.98 TeV, more probably 
3.75 TeV. We demonstrate that the same method 
can also regularize the Dirac monopole, which 
enhances the possibility to construct the Dirac 
monopole of mass of a few hundred meV in condensed 
matters. We discuss the physical implications of 
our result.
\end{abstract}

\keywords{Cho-Maison monopole, finite energy electroweak 
monopole, regularization of the electroweak monopole by 
electric charge renormalization, BPS electroweak monopole, 
BPS energy bound of electroweak monopole, lower bound of 
electroweak monopole mass, regularized Dirac monopole by 
vacuum polarization}
%%%%%%%%%%%%%%%%%%%%%%%%%%%%%%%%%%%
\maketitle

\section{Introduction}

Topology has played an essential role in physics. This is 
not accidental. All fundamental theories in physics are 
gauge theories which are described by the principal fiber 
bundle \cite{jmp75}. For example, the electrodynamics 
is described by the U(1) fiber bundle. And it has been 
well known that the principal fiber bundle has topological 
structure classified by the Chern classes. This tells that 
topology and physics can not be separated.

A best known example of such topological objects is 
the monopole. In 1931 Dirac has shown that, when 
the electromagnetic U(1) bundle becomes non-trivial,
we can have the Dirac monopole in electrodynamics which
obeys the famous charge quantization condition originating 
from the non-triviality of the U(1) bundle \cite{dirac}. 
The Dirac monopole has been generalized to non-Abelian 
monopoles \cite{wu,thooft,dokos,prl80}. In particular 
it has been shown that in the unification of the weak 
and electromagnetic interactions the Dirac monopole 
transforms to the electroweak monopole of the standard 
model \cite{plb97,yang}. 

In the course of the electroweak unification the Dirac
monopole changes it's character. First, it acquires 
the W-boson dressing and becomes a hybrid between 
the Dirac monopole and 'tHooft-Polyakov monopole. 
This is because the SU(2) part of the standard model 
naturally provides the $W$ boson dressing. Second, 
the magnetic charge becomes two times bigger. This 
is because the period of the electromagnetic U(1) 
becomes $4\pi$, not $2\pi$, in this unification. Third, 
unlike the Dirac monopole which is optional, this one 
must exist if the standard model is correct. This is because 
the unification inevitably makes the electromagnetic U(1) 
non-trivial \cite{plb97,yang}. 

This means that the discovery of this monopole, not 
the Higgs particle, should be interpreted as the final 
test (in fact the topological test) of the standard model.
Moreover, this monopole could play important roles 
in cosmology \cite{pta19}. In the early universe it could 
become the primordial black hole which could explain 
the dark matter, and become the seed of large scale 
structure of the universe. Most importantly, if discovered, 
this will become the first stable topological elementary 
particle in the history of physics.This makes the experimental detection of the electroweak monopole a most important 
issue after the discovery of the Higgs particle. For this 
reason the MoEDAL and ATLAS at LHC, the IceCube 
at the south pole, and similar detectors are actively 
searching for the monopole \cite{medal,atlas,icecu}.   

To detect the electroweak monopole, we have to know
the mass. This is a most important piece of information 
needed to detect the monopole. There was no way 
to predict the mass of the Dirac monopole theoretically, 
which has made the search for the monopole a blind 
search in the dark room. The electroweak monopole 
known as the Cho-Maison monopole is a hybrid between 
Dirac and 'tHooft-Polyakov \cite{plb97,yang}. As such 
it has a Dirac-type point singularity which makes the energy 
divergent, and classically we can not calculate the mass 
of the electroweak monopole. 

Fortunately we can estimate the mass. Intuitively the mass 
comes from exactly the same Higgs mechanism which 
makes the W-boson massive, except that here the coupling 
is magnetic. This implies that the monopole mass should 
be roughly $1/\alpha$ times bigger than the $W$ boson 
mass, of the order of 10 TeV. Moreover, one could argue 
that the quantum correction (the charge renormalization) 
might regularize the singularity and make the monopole 
energy finite. In fact we can obtain a finite energy 
electroweak monopole, replacing the hypercharge U(1) 
coupling by a running coupling, and show that 
the mass becomes around 7 TeV \cite{epjc15,mpla16}. 

But since the monopole mass is very important for us 
to detect the monopole a more precise estimate of 
the mass is needed. There have been two remarkable
reports in this direction. It has been argued that, making 
the hypercharge running coupling more realistic, one 
could put a constraint on the upper limit on the mass 
to be around 5.57 TeV \cite{ellis}. Moreover, modifying 
the standard model slightly one could show that 
the mass has the Bogomol'nyi-Prasard-Sommerfeld 
(BPS) bound 2.37 TeV \cite{bb}. 

The purpose of this paper is the following. First, we 
show that the singular Cho-Maison monopole can 
be regularized by the electromagnetic charge 
renormalization, the virtual electron-positron pair 
production, of the real electric charge. Second, we 
show that this regularization allows us to have a new 
BPS bound for the mass of the electroweak monopole 
given by 2.98 TeV, more realistically 3.75 TeV. Third, 
we demonstrate that the Dirac monopole can also 
be regularized by the same electric charge 
renormalization (i.e., the vacuum polarization). 

These results are important for the following reasons.
First, the regularization of the monopole with 
the electromagnetic permittivity assures that there 
is a realistic way to regularize the electroweak 
monopole. This enhances the possibility to find 
a finite energy electroweak monopole.

Second, the new BPS bound of the electroweak 
monopole mass is a very important information 
for the experiments searching for the electroweak 
monopole, in particular the MoEDAL and ATLAS 
experiments at LHC. This is because this implies 
that LHC may not be able to produce the monopole 
if the energy is less than 2.98 TeV.  

Third, the regularization of Dirac monopole is very 
interesting from theoretical point of view, because 
as far as we know no method to regularize the Dirac 
monopole has been known so far. Moreover, from 
the practical point of view this enhances 
the possibility to find the Dirac monopole greatly,
because this allows us to estimate mass of the Dirac 
monopole. Indeed, one of the reasons why the search 
for the Dirac monopole has not been successful is 
that it has been a blind search in the dark room, 
without any theoretical hint on the mass. Our
regularization changes this situation.

The paper is organized as follows. In Section II we review 
the known electroweak monopole and dyon solutions. 
In Section III we show how the electromagnetic charge 
screening could regularize the singular Cho-Maison
monopole and dyon. In Section IV we construct the BPS
electroweak monopole and set a new BPS bound for 
the monopole mass. In Section V we show that the same 
electromagnetic charge screening can regularize 
the Dirac monopole in electrodynamics and make 
the energy finite. Finally in the last section we discuss 
the physical implications of our results.

\section{Electroweak Monopole: A Review}

It has widely been believed that the topological structure 
of the standard model is fundamentally different from 
the Georgi-Glashow model, so that it has no monopole 
while the other one allows the 'tHooft-Polyakov monopole. 
This is not true. To see this we start from the (bosonic 
sector of) Weinberg-Salam model,
\begin{gather}
{\cal L} =-|{\cal D}_\mu \phi|^2 
-\frac{\lambda}{2}\big(|\phi|^2
-\frac{\mu^2}{\lambda}\big)^2-\frac{1}{4}\F_\mn^2
-\frac{1}{4}G_\mn^2, \nn \\
{\cal D}_\mu \phi =\big(\pd_\mu
-i\frac{g}{2} \vsig \cdot \A_\mu
-i\frac{g'}{2} B_\mu\big) \phi  \nn\\
=D_\mu \phi-i\frac{g'}{2} B_\mu \phi,  
\label{lag0}
\end{gather}
where $\phi$ is the Higgs doublet, $\A_\mu$, $\F_\mn$ 
and $G_\mn$, $B_\mu$ are the gauge fields of the SU(2) 
and hypercharge U(1), and $D_\mu$ is the covariant 
derivative of SU(2). 

Expressing $\phi$ with the Higgs field $\rho$ and unit 
doublet $\xi$ by
\begin{gather}
\phi = \dfrac{\rho}{\sqrt{2}}~\xi
~~~~(\xi^\dagger \xi = 1),
\end{gather}
we have
\begin{gather}
{\cal L}_{WS}=-\frac{1}{2} (\pd_\mu \rho)^2
- \frac{\rho^2}{2} |{\cal D}_\mu \xi |^2
-\frac{\lambda}{8}\big(\rho^2-\rho_0^2 \big)^2 \nn\\
-\frac14 \F_\mn^2 -\frac14 G_\mn^2,
\end{gather}
where $\rho_0=\sqrt{2\mu^2/\lambda}$ is the vacuum 
expectation value of the Higgs field. Notice that 
the hypercharge U(1) gauge interaction makes $\xi$  
a $CP^1$ field \cite{plb97}. which can naturally admit 
the $\pi_2(S^2)$ monopole topology. 

To simplify this further we need the Abelian decomposition 
of the standard model. Start from the SU(2) gauge field 
$\A_\mu$ and let $(\n_1,\n_2,\n_3=\n)$ be an arbitrary 
right-handed orthonormal $SU(2)$ basis. Choose $\n$ 
to be the Abelian direction at each space-time point, 
and project out the restricted potential $\hA_\mu$ 
imposing the isometry condition \cite{prd80,prl81}
\begin{gather}
D_\mu \hn=0,  \nn\\
\A_\mu\rightarrow \hA_\mu =\tA_\mu +\tC_\mu, \nn\\
\tA_\mu= A_\mu \n~~(A_\mu=\n \cdot \A_\mu),   
~~~\tC_\mu=-\frac{1}{g} \n\times \pd_\mu \n.
\label{icon}
\end{gather}
The restricted potential has a dual structure, made of 
two potentials $\tA_\mu$ and $\tC_\mu$ which can also 
be described by two Abelian potentials $A_\mu$ and 
$C_\mu$,
\begin{gather}
\hF_\mn= \pd_\mu \hA_\nu-\pd_\nu \hA_\mu
+ g \hA_\mu \times \hA_\nu =F_\mn' \n, \nn \\
F'_\mn=F_\mn + H_\mn
= \pd_\mu A'_\nu-\pd_\nu A'_\mu,  \nn\\
F_\mn =\pd_\mu A_\nu-\pd_\nu A_\mu, \nn\\
H_\mn = -\frac1g \n \cdot (\pd_\mu \n \times\pd_\nu \n)
=\pd_\mu C_\nu-\pd_\nu C_\mu,  \nn\\
\n=-\xi^\dagger \vsig \xi,   \nn\\
C_\mu = -\frac{2i}{g} \xi^\dagger \pd_\mu \xi
=-\frac1g \n_1\cdot \pd_\mu \n_2,   \nn\\
A_\mu' = A_\mu+ C_\mu.
\end{gather}
Notice that the potential $C_\mu$ is determined uniquely 
up to the $U(1)$ gauge freedom which leaves $\n$ 
invariant. With
\begin{gather}
\xi =\exp (-i \gamma) \left(\begin{array}{cc}
\sin \frac{\alpha}{2}~\exp (-i \beta) \\
- \cos \frac{\alpha}{2} \end{array} \right), 
\label{xi}
\end{gather}
we have
\begin{gather}
\n=\left(\begin{array}{ccc} 
\sin \alpha \cos \beta \\
\sin \alpha \sin \beta \\  
\cos \alpha  \end{array} \right),
~~~C_\mu=-\frac1g (1-\cos \alpha) \pd_\mu \beta.
\label{monp}
\end{gather}
So when $\n=\hr$, the potential $C_\mu$ describes 
the Dirac monopole and $\tC_\mu$ describes 
the Wu-Yang monopole \cite{wu,prl80}. 

With this we obtain the gauge independent Abelian 
decomposition of the $SU(2)$ gauge field adding 
the gauge covariant valence part $\W_\mu$ which 
was excluded by the isometry \cite{prd80,prl81}
\begin{gather} 
\A_\mu = \hA_\mu + \W_\mu, 
~~~\W_\mu =W^1_\mu ~\n_1 + W^2_\mu ~\n_2,    \nn\\
\F_\mn=\hF_\mn + \hD _\mu \W_\nu - \hD_\nu
\W_\mu + g\W_\mu \times \W_\nu,   \nn\\
\hD_\mu=\pd_\mu+g \hA_\mu \times.
\label{cdec}
\end{gather}
This is the gauge independent Abelian decomposition of 
the SU(2) gauge theory known as the Cho decomposition, Cho-Duan-Ge (CDG) decomposition, or Cho-Faddeev-Niemi (CFN) decomposition \cite{fadd,shab,zucc,kondo}. Notice 
that once $\n$ is chosen, the decomposition follows automatically, regardless of the choice of gauge. 

With this we can abelianize the Weinberg-Salam theory.
Defining the electromagnetic field, $W$ boson, and 
$Z$ boson by
\begin{gather}
\left( \begin{array}{cc} A_\mu^{\rm (em)} \\  Z_{\mu}
\end{array} \right)
=\frac{1}{\sqrt{g^2 + g'^2}} \left(\begin{array}{cc} g & g' \\
-g' & g \end{array} \right)
\left( \begin{array}{cc} B_{\mu} \\ A'_{\mu}
\end{array} \right)  \nn\\
= \left(\begin{array}{cc}
\cos\theta_{\rm w} & \sin\theta_{\rm w} \\
-\sin\theta_{\rm w} & \cos\theta_{\rm w}
\end{array} \right)
\left( \begin{array}{cc} B_{\mu} \\ A_\mu'
\end{array} \right),   \nn\\
W_\mu =\frac{1}{\sqrt{2}} (W^1_\mu + i W^2_\mu), 
\label{mix}
\end{gather}
we have 
\begin{gather}
|D_\mu \xi|^2 =\frac{g^2}{4} ({A'}_\mu^2 
+ \W_\mu^2),  \nn\\
|{\cal D}_\mu \xi|^2=|D_\mu \xi|^2 
+ig' B_\mu \xi^\dagger D_\mu \xi 
+\frac{g'^2}{4} B_\mu^2  \nn\\
=\frac14 \W_\mu^2 +\frac{g^2+g'^2}{4} Z_\mu^2,
\label{id1}
\end{gather}
so that
\begin{gather}
{\cal L}=-\frac{1}{2} (\partial_\mu \rho)^2
-\frac{\lambda}{8}\big(\rho^2 -\rho_0^2 \big)^2
- \frac{1}{4} {F'}_\mn^2  \nn\\
-\frac14 G_\mn^2
-\frac{1}{2} |D'_\mu W_\nu-D'_\nu W_\mu|^2 \nn\\
- \frac{g^2}{4} \rho^2 W_\mu^* W_\mu  
-\frac{g^2+g'^2}{8} \rho^2 Z_\mu^2  \nn\\
+ ig F'_\mn W_\mu^*W_\nu 
+ \frac{g^2}{4}(W_\mu^* W_\nu - W_\nu^* W_\mu)^2  \nn\\
= -\frac12 (\pd_\mu \rho)^2 
-\frac{\lambda}{8}\big(\rho^2-\rho_0^2 \big)^2 
-\frac14 {F_\mn^{\rm (em)}}^2 \nn\\
-\frac12 \big|(D_\mu^{\rm (em)} 
+ie\frac{g}{g'} Z_\mu) W_\nu -(D_\nu^{\rm (em)} 
+ie\frac{g}{g'} Z_\nu) W_\mu)\big|^2  \nn\\
-\frac14 Z_\mn^2 -\frac{g^2}{4}\rho^2 W_\mu^*W_\mu
-\frac{g^2+g'^2}{8} \rho^2 Z_\mu^2   \nn\\
+ie (F_\mn^{\rm (em)} 
+\frac{g}{g'}  Z_\mn) W_\mu^* W_\nu  \nn\\
+ \frac{g^2}{4}(W_\mu^* W_\nu - W_\nu^* W_\mu)^2,
\label{lag1}
\end{gather}
where $D'_\mu=\pd_\mu+igA'_\mu$,
$D_\mu^{\rm (em)}=\pd_\mu+ieA_\mu^{\rm (em)}$
and $e$ is the electric charge
\begin{gather}
e=\frac{gg'}{\sqrt{g^2+g'^2}}=g\sin\theta_{\rm w}
=g'\cos\theta_{\rm w}.
\end{gather}
We emphasize that this is not the Weinberg-Salam 
Lagrangian in the unitary gauge. This is a gauge 
independent expression.

The abelianization sheds a new light on the standard model. 
First of all, here the Higgs doublet disappears completely.
Moreover, the $W$ boson and $Z$ boson acquire the mass 
when $\rho$ has the non-vanishing vacuum expectation 
value, without any "spontaneous" symmetry breaking 
by the Higgs doublet. 

As importantly this clarifies the topological structure of 
the standard model. It has been asserted that the standard 
model has no monopole topology \cite{col,vach}. The basis 
of this ``no-go theorem" is that, unlike the Higgs triplet 
in the Georgi-Glashow model, the Higgs doublet breaks 
the SU(2) symmetry completely. The above exercise, 
however, shows that this is wrong. First, the Higgs doublet 
disappears completely in this Abelianization. Second, 
in the absence of the weak bosons (\ref{lag1}) reduces 
to electrodynamics, which admits the Dirac monopole. 
More importantly, in this  Abelianization the electromagnetic 
U(1) becomes non-trivial. This is evident from the fact that 
$A'_\mu$ contains the monopole potential $C_\mu$. This 
confirms that, unlike in QED where the monopole becomes 
optional, the standard model must have the monopole. 

In fact, this monopole topology can be traced back to  
the original Lagrangian (\ref{lag0}). With the separation 
of the overall U(1) phase, (\ref{xi}) shows that $\xi$ 
becomes a $CP^1$ field which has the $S^2$ topology. 
Moreover, $\n$ defined by $\xi$ has the $S^2$ topology. 
So (\ref{lag0}) has the $\pi_2(S^2)$ topology which 
describes the monopole. 

To construct the monopole we choose the ansatz in 
the spherical coordinates $(t,r,\theta,\varphi)$
\begin{gather}
\rho=\rho(r),~~~\xi =i\left(\begin{array}{cc}
\sin \frac{\theta}{2}~e^{-i\varphi} \\
- \cos \frac{\theta}{2} \end{array} \right), \nn\\
\n=- \xi^\dagger \vsig \xi=\hr ,  
~~~C_\mu= -\frac1g (1-\cos \theta) \pd_\mu \varphi,  \nn\\
\A_\mu= \frac{1}{g} A(r) \pd_\mu t~\hr
+\frac{1}{g}(f(r)-1)~\hr \times \pd_\mu \hr, \nn\\
B_{\mu} =\frac{1}{g'} B(r) \pd_\mu t
-\frac{1}{g'}(1-\cos\theta) \pd_\mu \varphi.
\label{ans0}
\end{gather}
Notice that the apparent string singularity along the negative 
$z$-axis in $\xi$ and $B_{\mu}$ is a pure gauge artefact 
which can easily be removed making $U(1)$ non-trivial. 
In terms of the physical fields the ansatz is expressed by
\begin{gather}
A_\mu^{\rm (em)}=e\big(\frac{A(r)}{g^2}
+\frac{B(r)}{g'^2} \big)
\pd_\mu t -\frac1e (1-\cos\theta)\pd_\mu \varphi,  \nn\\
W_\mu= \frac{i}{g} \frac{f(r)}{\sqrt 2}e^{i\varphi}
(\pd_\mu \theta +i \sin\theta \pd_\mu \varphi), \nn\\
Z_\mu= \frac{e}{gg'}\big(A(r)-B(r) \big)\pd_\mu t,
\label{ans1}
\end{gather}
which assures that the ansatz is for a dyon. 

\begin{figure}
\includegraphics[height=4.5cm, width=8cm]{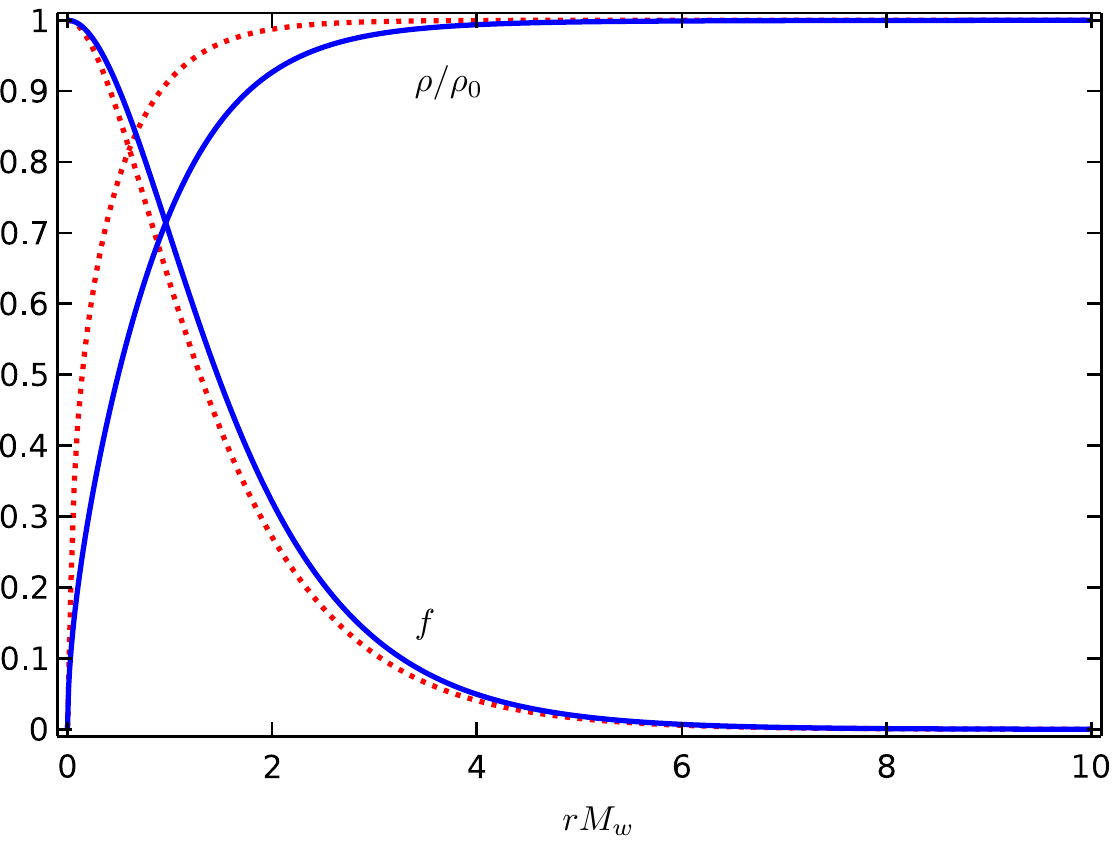}
\caption{\label{cmm} The electroweak monopole solutions. 
The red curves represent the singular Cho-Maison monopole
obtained by (\ref{cdeq}) and the blue curves represent  
the finite energy electroweak monopole obtained by (\ref{fedeq}) with $A=B=0$.}
\end{figure}

With the ansatz we have the dyon equations of motion
\begin{gather}
\ddot \rho+\frac{2}{r}\dot \rho -\frac{1}{2r^2} f^2 \rho
=-\frac14 (B-A)^2 \rho 
+\frac{\lambda}{2} \big(\rho^2-\rho_0^2 \big) \rho, \nn \\
\ddot f - \frac{1}{r^2} (f^2 -1) f
=\big(\frac{g^2}{4} \rho^2-A^2 \big)~f,  \nn \\
\ddot A +\frac{2}{r} \dot A - \frac{2}{r^2} f^2 A
=\frac{g^2}{4}(A-B)~\rho^2,  \nn\\
\ddot B+\frac{2}{r}\dot B =\frac{g'^2}{4}(B-A)~\rho^2,
\label{cdeq}
\end{gather}
which has the energy 
\begin{gather}
E=4\pi \int_0^\infty dr \bigg\{ \frac{1}{2g'^2 r^2}
+\frac12 (r\dot\rho)^2
+\frac{\lambda}{8} r^2 \big(\rho^2-\rho_0^2 \big)^2 \nn\\
+\frac1{g^2} \big(\dot f^2 + \frac{(f^2-1)^2}{2r^2} 
+ f^2 A^2 \big) +\frac14 f^2\rho^2   \nn\\
+\frac{r^2}{2} \big(\frac{\dot A^2}{g^2}
+\frac{\dot B^2}{g'^2}  \big)   
+\frac{r^2}{8} (B-A)^2 \rho^2 \bigg\}.
\label{cde}
\end{gather}
Obviously (\ref{cdeq}) has the singular monopole 
solution which describes the point monopole in 
Weinberg-Salam model
\begin{gather}
f=0,~~~\rho=\rho_0,~~~A=B=0,  \nn\\
A_\mu^{\rm (em)} = -\frac{1}{e}(1-\cos \theta) 
\pd_\mu \varphi,
\label{cms}
\end{gather}
which has the magnetic charge $4\pi/e$ (not $2\pi/e$).

\begin{figure}
\includegraphics[height=4.5cm, width=8cm]{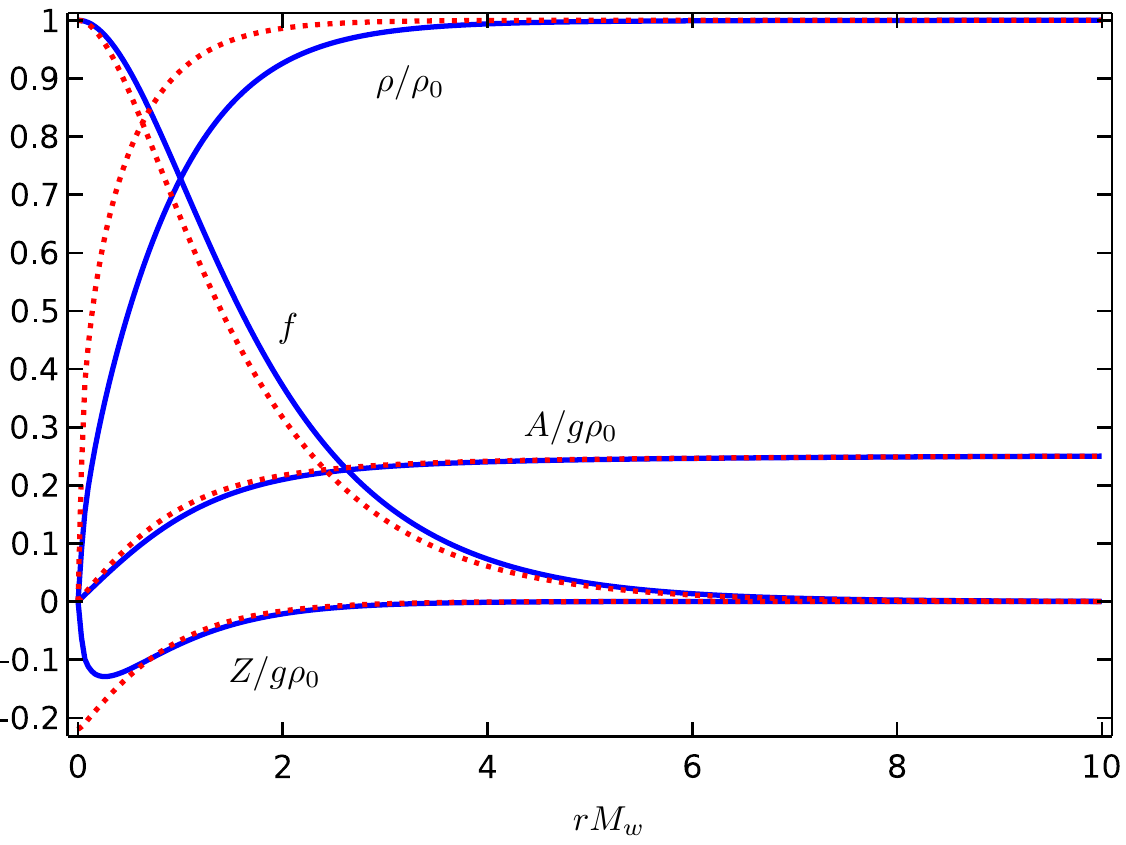}
\caption{\label{cmd} The electroweak dyon solution.
The red curves represent the singular Cho-Maison dyon 
and the blue curves represent the finite energy dyon 
obtained by (\ref{fedeq}). Here we have put 
$A_0=M_W/2$ and $Z=A-B$.} 
\end{figure}

Integrating (\ref{cdeq}) with $A=B=0$ and with 
the boundary condition
\begin{gather}
\rho(0)=0,~~~f(0)=1,
~~~\rho(\infty)=\rho_0,~~~f(\infty)=0,
\label{bcm}
\end{gather}
we obtain the singular (Cho-Maison) monopole solution 
dressed by the $W$ boson and Higgs field shown in 
Fig. \ref{cmm} by the red curves. Moreover, with 
the boundary condition
\begin{gather}
\rho(0)=0,~~f(0)=1,~~A(0)=0,~~B(0)=b_0, \nn\\
\rho(\infty)=\rho_0,~f(\infty)=0,
~A(\infty)=B(\infty)=A_0,
\label{bcd}
\end{gather}
we can obtain the singular (Cho-Maison) dyon solution which 
has the extra electric charges $q_e=(4\pi A_1)/e$. This is 
shown by the red curves in Fig. \ref{cmd}.

The electroweak monopole can be viewed as a hybrid
between the Dirac monopole and the 'tHooft-Polyakov
monopole, so that it has a U(1) point singularity at
the center even though the SU(2) part is completely
regular. 

We can regularize the point singularity with the quantum correction at short distance, replacing the coupling 
constant $g'$ of the hypercharge U(1) to an effective 
coupling which diverges at the origin \cite{epjc15,mpla16}. 
To show this, we modify the Lagrangian (\ref{lag1}) 
introducing a non-trivial hypercharge U(1) permittivity 
$\eps (\rho)$ which depends on $\rho$,
\begin{gather}
{\cal L}=-\frac{1}{2} (\partial_\mu \rho)^2
-\frac{\lambda}{8}\big(\rho^2 -\rho_0^2 \big)^2
- \frac{1}{4} {F'}_\mn^2  \nn\\
-\frac14  \eps(\rho)~G_\mn^2
-\frac{1}{2} |D'_\mu W_\nu-D'_\nu W_\mu|^2 \nn\\
- \frac{g^2}{4} {\rho}^2 W_\mu^* W_\mu  
-\frac18 \rho^2 (gA'_\mu-g'B_\mu)^2  \nn\\
+ ig F'_\mn W_\mu^*W_\nu 
+ \frac{g^2}{4}(W_\mu^* W_\nu - W_\nu^* W_\mu)^2.
\label{effl1}
\end{gather}
The effective Lagrangian retains the $SU(2)\times U(1)$ 
gauge symmetry. Moreover, when $\eps$ approaches 
to one asymptotically, it reproduces the standard model.
But $\eps$ effectively changes the $U(1)_Y$ gauge coupling 
$g'$ to the ``running" coupling $\bg'=g' /\sqrt{\eps}$. 
This is because with the rescaling of $B_\mu$ to $B_\mu/g'$, 
$g'$ changes to $g' /\sqrt{\eps}$. So, by making $\bg'$ 
infinite at the origin, we can regularize the Cho-Maison 
monopole \cite{epjc15,mpla16}.

From this we have the modified dyon equation 
\begin{gather}
\ddot \rho + \frac{2}{r}\dot \rho-\frac{f^2}{2r^2} \rho 
=\frac{\lambda}{2} (\rho^2- \rho_0^2) \rho
-\frac{1}{4} (A-B)^2 \rho  \nn\\
+\frac{\eps'}{2 g'^2}\Big(\frac{1}{r^4}-\dot{B}^2 \Big),  \nn\\
\ddot{f}-\frac{f^2-1}{r^2}f
=\big(\frac{g^2}{4} \rho^2 - A^2\big)f, \nn\\
\ddot{A}+\frac{2}{r}\dot{A}-\frac{2f^2}{r^2}A
=\frac{g^2}{4} \rho^2(A-B), \nn \\
\ddot{B} + 2\big(\frac{1}{r}
+\frac{\eps'}{2 \eps} \dot \rho \big) \dot{B}  
=-\frac{g'^2}{4 \epsilon} \rho^2 (A-B),
\label{fedeq}
\end{gather}
where $\eps' = d\eps/d\rho$. This has the energy 
\begin{gather}
E=4\pi \int_0^\infty dr \bigg\{\frac{\eps}{2g'^2 r^2} 
+\frac12 (r\dot\rho)^2
+\frac{\lambda}{8} r^2 \big(\rho^2-\rho_0^2 \big)^2 \nn\\
+\frac1{g^2} \big(\dot f^2 + \frac{(f^2-1)^2}{2r^2} 
+ f^2 A^2 \big) +\frac14 f^2\rho^2  \nn\\
+\frac{(r\dot A)^2}{2g^2}+\frac{\eps (r\dot B)^2}{2g'^2}
+\frac{r^2}{8} (A-B)^2 \rho^2 \bigg\}.
\label{fede}
\end{gather}
To have a regular solution we let near the origin 
\begin{gather} 
\rho(r) = r^\delta (h_0+h_1 r+...),   \nn\\
\beps =\Big(\frac{\rho}{\rho_0}\Big)^n  
\big[c_0+c_1 (\frac{\rho}{\rho_0})+...\big],  
\label{eps}
\end{gather}
and find from the equation of motion that we need 
\begin{gather}
\delta={\frac{\sqrt{3}-1}{2}},
~~~n > \frac{2}{\delta} =2(\sqrt 3+1)
\simeq 7.46,
\label{rcon1}
\end{gather}
or
\begin{gather}
\delta={\frac{2}{n-2}},
~~~2< n \leq 2(\sqrt 3+1).
\label{rcon2}
\end{gather}
This assures that when $n > 2$ we have finite energy 
monopole and dyon.

We can integrate (\ref{fedeq}) with 
$\eps=(\rho/\rho_0)^n$. The regularized monopole 
and dyon solutions with $n=6$ are shown in 
Fig. \ref{cmm} and Fig. \ref{cmd} by the blue curves. 
Notice that asymptotically the regularized solutions 
look very much like the singular solutions, except that for 
the finite energy dyon solution $Z$ becomes zero at 
the origin. With $n=6$ the monopole energy becomes
\begin{gather}
E \simeq 0.65 \times \frac{4\pi}{e^2} M_W 
\simeq 7.20 ~{\rm TeV}.
\end{gather}
This confirms that the ultraviolet regularization of 
the Cho-Maison dyon is indeed possible. 

Of course, we could choose different $\eps$ to obtain
different solution. Indeed choosing the realistic $\eps$ 
which can fit the Higgs to two photon decay data better, 
one could argue that the mass of the electroweak 
monopole may not be larger than 5.57 TeV \cite{ellis}.   

One can have the analytic extension of the Cho-Maison 
monopole \cite{epjc15}. To see this notice that in 
the absence of the $Z$ boson (\ref{lag1}) reduces to
\begin{gather}
{\cal L}= -\frac12 (\pd_\mu \rho)^2 
-\frac{\lambda}{8}\big(\rho^2-\rho_0^2 \big)^2 
-\frac14 {F_\mn^{\rm (em)}}^2 \nn\\
-\frac12 |(D_\mu^{\rm (em)} W_\nu 
-D_\nu^{\rm (em)} W_\mu)|^2 
+ie F_\mn^{\rm (em)} W_\mu^* W_\nu    \nn\\
-e^2 \rho^2 W_\mu^*W_\mu  
+ \frac{e^2}{4}(W_\mu^* W_\nu - W_\nu^* W_\mu)^2 
+\delta {\cal L},   \nn\\
\delta {\cal L}=\big(e^2- \frac{g^2}{4} \big)
\rho^2 W_\mu^*W_\mu \nn\\
-\frac{e^2-g^2}{4} (W_\mu^* W_\nu - W_\nu^* W_\mu)^2.
\label{lag2}
\end{gather}
Notice that when $\delta {\cal L}=0$ this reduces 
to the Georgi-Glashow Lagrangian which describes 
the spontaneously broken SU(2) gauge theory \cite{epjc15}.

\begin{figure}
\includegraphics[height=4.5cm, width=8cm]{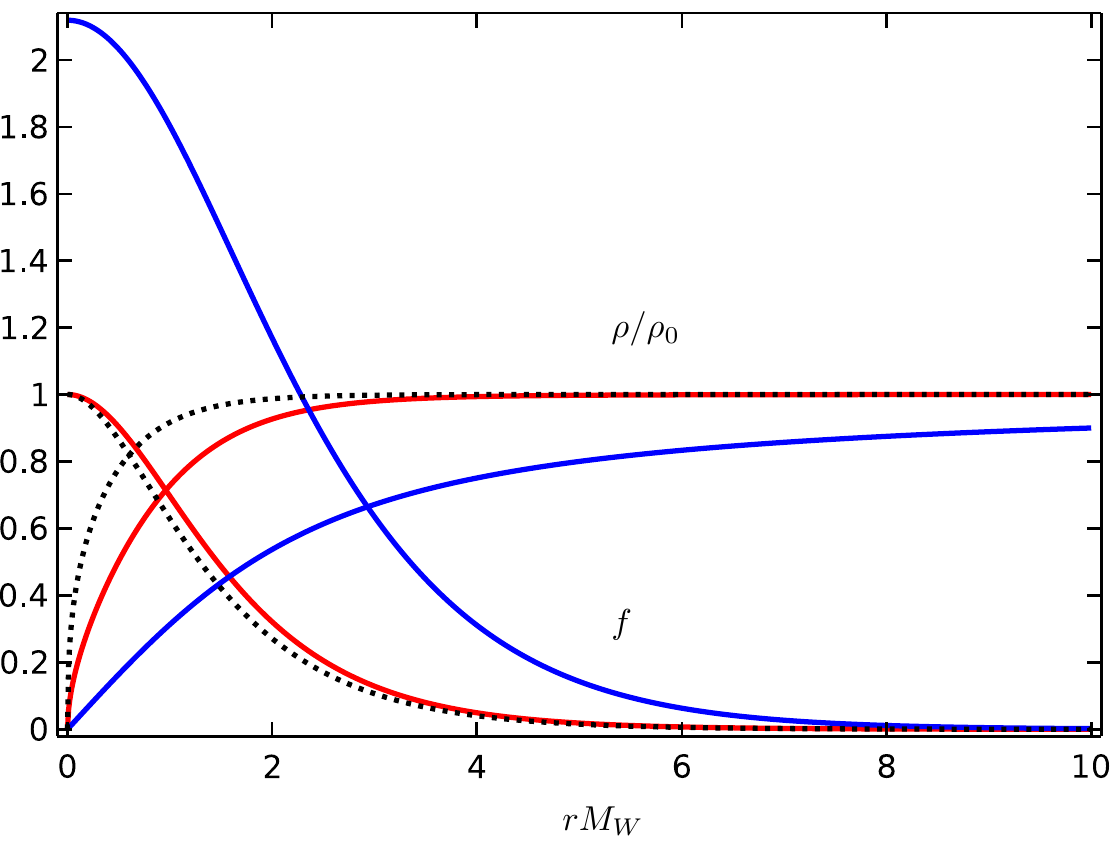}
\caption{\label{acm} The analytic electroweak 
monopole solution. The blue curves represent 
the analytic monopole given by (\ref{cmam}) and 
the red curves represent the finite energy electroweak 
monopole obtained by (\ref{fedeq}). The Cho-Maison 
monopole is shown in dotted curves for comparison.} 
\end{figure}

With the monopole ansatz this has the energy density in
the BPS limit (i.e., in the limit $\lambda$ vanishes)
\begin{gather}
E =\frac{2\pi}{g^2}\int_0^\infty dr 
\bigg\{g^2 \Big(r \dot \rho
- \frac{1}{e r} (\frac{e^2}{g^2} f^2-1) \Big)^2  \nn\\
+2 \big(\dot f - e \rho f \big)^2 
+\frac{2 g^2}{e} \dot \rho~(\frac{e^2}{g^2} f^2-1) 
+ 4e \rho \dot f f \nn\\
+(\frac{g^2}{2}-2e^2)\frac{\rho^2f^2}{r^2}
+\frac{g^2-e^2}{g^2 r^4}f^4  \bigg\}.
\label{bpse}
\end{gather} 
So, when the following BPS equation holds 
\begin{gather}
\dot{\rho} -\frac{1}{e r^2}\big(\frac{e^2}{g^2}f^2-1 \big)=0,
~~~\dot{f} -e\rho f=0.
\label{bpseq}
\end{gather}
the monopole has the analytic solution \cite{epjc15} 
\begin{gather}
\rho=\rho_0\coth(e\rho_0r)-\frac{1}{er}, 
~~~f= \frac{g\rho_0 r}{\sinh(e\rho_0r)}.
\label{cmam}
\end{gather}
The BPS electroweak monopole is shown by the blue curves 
in Fig. \ref{acm}. Notice that in this solution we have 
$f(0)=g/e$.

With (\ref{bpseq}) the energy has the bound
\begin{gather}
E \geq \frac{4\pi}{g^2}\int_0^\infty dr 
\Big\{\frac{g^2}{e^2 r^2} (\frac{e^2}{g^2} f^2-1)^2 
+ 2 e^2 \rho^2 f^2 \nn\\
+(\frac{g^2}{4}-e^2)\frac{\rho^2f^2}{r^2}
+\frac{e^2}{2g'^2r^4}f^4  \Big\}.
\label{bpseb}
\end{gather} 
Notice that the last two terms come from $\delta \cL$.
So, if we neglect $\delta \cL$ in the standard model, 
the model has the analytic solution similar to 
the Prasard-Sommerfeld solution in Georgi-Glashow 
model which could set the BPS bound for the monopole 
mass. Unfortunately the last term in (\ref{bpseb}) 
is divergent, so that the standard model as it is has 
no BPS bound monopole solution. As we will see, however,
we can have the BPS electroweak monopole when we 
modify the standard model.

\section{Regularization of Electroweak Monopole with 
Electromagnetic Permittivity}

The above discussion shows that the quantum correction
(the renormalization of the hypercharge coupling) could regularize the electroweak monopole and make 
the energy finite. Now we show that the renormalization 
of the real electric charge, can also regularize 
the Cho-Maison monopole. 

This is because the point singularity of $B_\mu$ 
translates to the point singularity of $A_\mu^{\rm (em)}$. 
To demonstrate this, consider the following effective 
Lagrangian of the standard model
\begin{gather}
{\cal \bar L}= -\frac12 (\pd_\mu \rho)^2 
-\frac{\lambda}{8}\big(\rho^2-\rho_0^2 \big)^2 
-\frac14 \beps(\rho) {F_\mn^{\rm (em)}}^2 \nn\\
-\frac12 \big|(D_\mu^{\rm (em)} 
+ie\frac{g}{g'} Z_\mu) W_\nu -(D_\nu^{\rm (em)} 
+ie\frac{g}{g'} Z_\nu) W_\mu)\big|^2  \nn\\
-\frac14 Z_\mn^2 -\frac{g^2}{4}\rho^2 W_\mu^*W_\mu
-\frac{g^2+g'^2}{8} \rho^2 Z_\mu^2   \nn\\
+ie (F_\mn^{\rm (em)} 
+\frac{g}{g'}  Z_\mn) W_\mu^* W_\nu   \nn\\
+ \frac{g^2}{4}(W_\mu^* W_\nu - W_\nu^* W_\mu)^2,
\label{lag3}
\end{gather}
where $\beps$ is the real non-vacuum electromagnetic 
permittivity. 

With the ansatz (\ref{ans1}) we have the following dyon
equations of motion
\begin{gather}
\ddot \rho + \frac{2}{r}\dot \rho-\frac{f^2}{2r^2} \rho
=\frac{\lambda}{2} (\rho^2- \rho_0^2) \rho  
-\frac{1}{4} (B-A)^2 \rho    \nn\\
+\frac{\beps'}{2}\Big(\frac{1}{e^2 r^4}
-e^2(\frac{\dot A}{g^2}+\frac{\dot B}{g'^2})^2 \Big),  \nn\\
\ddot{f}-\frac{f^2-1}{r^2}f
=\big(\frac{g^2}{4} \rho^2 - A^2\big) f, \nn\\
\ddot A + \frac{2}{r} \dot A
+e^2 \frac{\beps'}{\beps}\dot \rho \big(\frac{\dot A}{g^2}
+\frac{\dot B}{g'^2}\big)   
-\frac{2 e^2}{g^2} \big(\frac{g^2}{g'^2} +\frac{1}{\beps}\big) 
\frac{f^2}{r^2} A  \nn\\
=-\frac{g^2}{4} \rho^2 (B-A),  \nn\\
\ddot B +\frac{2}{r}\dot B
+e^2 \frac{\beps'}{\beps} \dot \rho \big(\frac{\dot A}{g^2}
+\frac{\dot B}{g'^2}\big)
+\frac{2e^2}{g^2} \big(1
-\frac{1}{\beps} \big) \frac{f^2}{r^2} A \nn\\
=\frac{g'^2}{4}(B-A) \rho^2.
\label{fedeq2}
\end{gather}
where $\beps'=d \beps/d\rho$. To integrate it out and 
find a finite energy solution we have to choose a proper 
boundary condition.

\begin{figure}
\includegraphics[height=4.5cm, width=8cm]{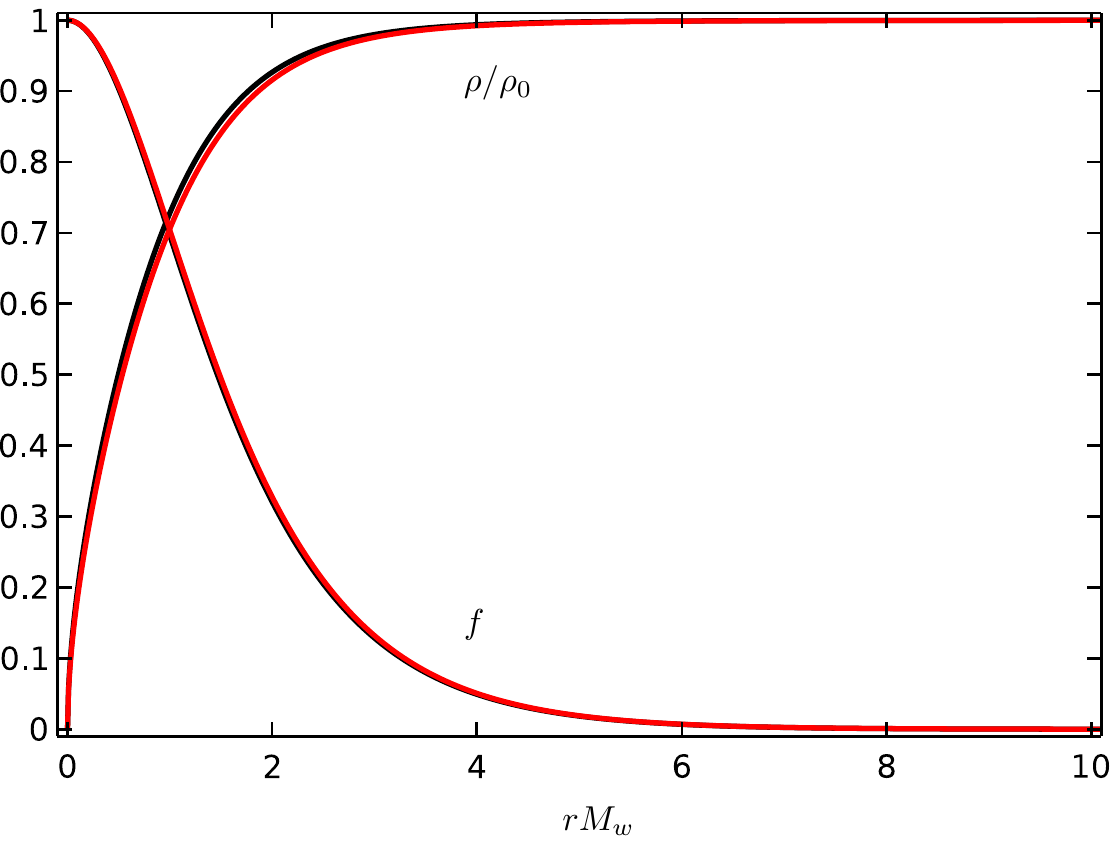}
\caption{\label{fem} The finite energy electroweak monopole 
solution regularized by the electromagnetic permittivity 
with W boson and Higgs scalar dressing. The red curve 
represents the regularized monopole solution with 
the non-trivial permittivity $\beps_1=(\rho/\rho_0)^6$. 
For comparison we plot the monopole solution regularized 
by the hypercharge renormalization  with the black 
curve (shown in Fig. \ref{cmm}).}
\end{figure}

To find a proper boundary condition notice that
the Lagrangian gives us the following dyon energy
\begin{gather}
E=4\pi \int_0^\infty dr \bigg\{\frac{\beps}{2e^2 r^2} 
+\frac12 (r\dot \rho)^2
+\frac{\lambda r^2}{8}\big(\rho^2-\rho_0^2 \big)^2 \nn\\
+\frac1{g^2} \big(\dot f^2 + \frac{f^2(f^2-2)}{2r^2} \big) 
+\frac14 f^2 \rho^2 + \frac{f^2 A^2}{g^2}  \nn\\
+\frac{r^2}{8}(A-B)^2\rho^2
+\frac{r^2(\dot A-\dot B)^2}{2(g^2+g'^2)}  \nn\\
+\frac{\beps~e^2 r^2}{2} 
\big(\frac{\dot A}{g^2} +\frac{\dot B}{g'^2} \big)^2 \bigg\}.
\label{fede1}
\end{gather}
This has two potentially divergent terms near the origin, 
the first and fifth terms. Assuming (\ref{eps}) we can 
make the first term finite. To make the fifth term finite 
we might require$f(0)=0$ or $f(0)=2$. But we find that 
this condition does not yield a finite energy solution. 

A correct way to regularize the solution is to combine 
the two divergent terms and make it finite, since both 
have the same $O(r^{-2})$ divergence at the origin. 
So, we let
\begin{gather}
\beps =\beps_0 +\beps_1,
~~~\beps_0= \frac{g'^2}{g^2+g'^2} ,  
~~~\beps_1 \simeq \Big(\frac{\rho}{\rho_0}\Big)^n,  
\label{epbc}
\end{gather}  
and find that the dyon energy is expressed by
\begin{gather}
E=4\pi \int_0^\infty dr \bigg\{\frac{1}{2e^2 r^2}
\Big(\beps_0 (f^2-1)^2+ \beps_1 \Big)
+\frac12 (r\dot \rho)^2   \nn\\
+\frac{\lambda r^2}{8}\big(\rho^2-\rho_0^2 \big)^2
+\frac1{g^2} \dot f^2 +\frac14 f^2 \rho^2 
+ \frac{f^2 A^2}{g^2}  \nn\\
+\frac{r^2}{8}(A-B)^2\rho^2
+\frac{r^2(\dot A-\dot B)^2}{2(g^2+g'^2)}  \nn\\
+\frac{\beps~e^2 r^2}{2} \big(\frac{\dot A}{g^2}
+\frac{\dot B}{g'^2} \big)^2  \bigg\}.
\label{fede2}
\end{gather}
This tells that we can make the energy finite imposing 
the boundary condition $f(0)=1$. In other words we 
can regularize the monopole with the real electromagnetic 
permittivity $\beps$ making $\beps(0)$ finite, with 
the same boundary condition for $f$ as before. This is remarkable.
 
\begin{figure}
\includegraphics[height=4.5cm, width=8cm]{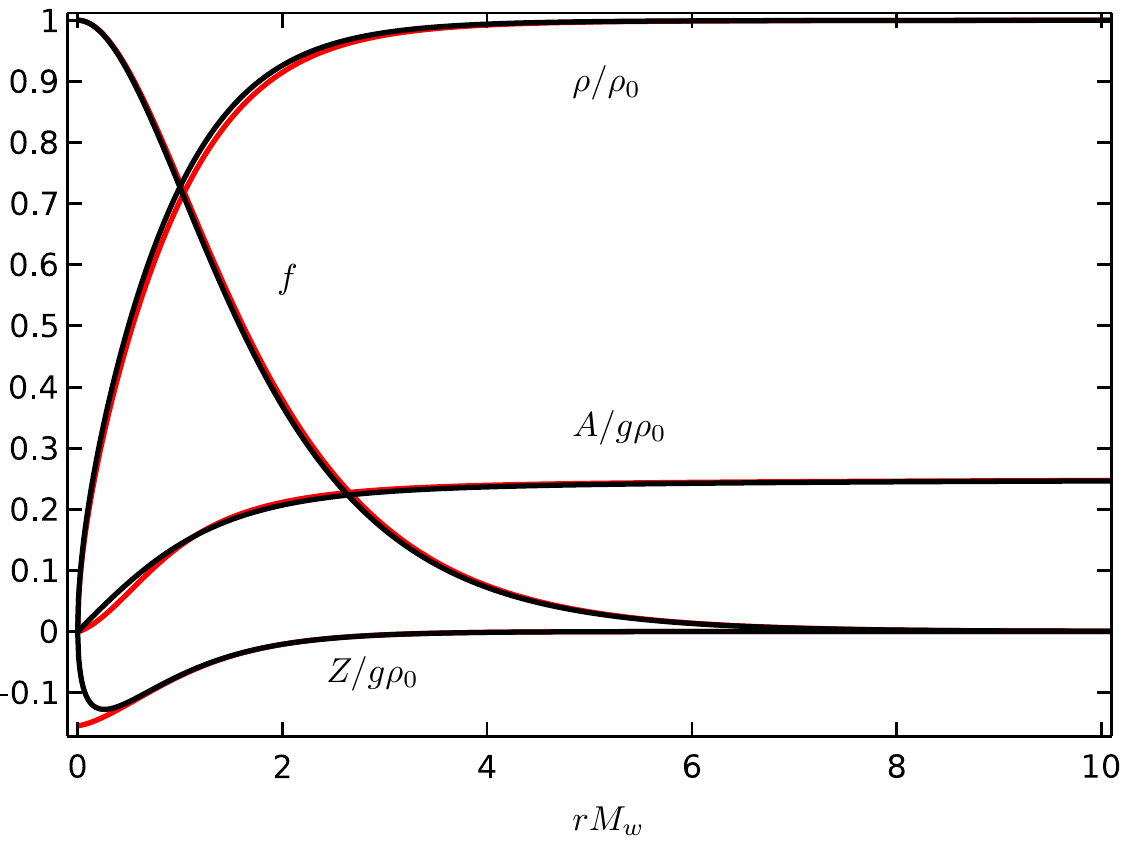}
\caption{\label{fed} The finite energy dyon solution
(the red curves) regularized by the real electromagnetic 
permittivity $\beps_1=(\rho/\rho_0)^6$ with the W boson,
Z boson, and Higgs field dressing. For comparison we 
plot the finite energy dyon solution regularized by 
the hypercharge renormalization (shown in Fig. \ref{cmd}) 
by the black curves.}
\end{figure} 
 
To see that this regularization works, consider the monopole 
solution first. In this case we can integrate (\ref{fedeq2}) 
with $\beps_1=(\rho/\rho_0)^n$, $A=B=0$, and with 
the boundary condition $f(0)=1$, and obtain the finite 
energy monopole solution (with $n>2$). The monopole
solution with $n=6$ is shown in Fig. \ref{fem}.
We could generalize the monopole solution to dyon 
by solving (\ref{fedeq2}). The solution with $n=6$ is 
shown in Fig. \ref{fed}. Remarkably the solutions look 
almost identical to the solutions shown in Fig. \ref{cmm}
and Fig. \ref{cmd}. But the difference is that here $\beps$ 
has a non-vanishing value $\beps_0$ at the origin. 

Of course, the energy depends on $n$, and we can plot 
the monopole energy in terms of $n$. This is shown in 
Fig. \ref{mevn} in red dots. For $n=6$ the energy is given 
by 7.96 TeV. For comparison we plot the monopole energy 
regularized by $\eps$ (by the hypercharge renormalization) 
in blue dots. Notice that the energy approaches to 
the asymptotic value roughly given by 3.75 TeV shown 
in black dotted line.  

This confirms that we can indeed regularize the electroweak 
monopole and dyon solutions with the real electromagnetic 
permittivity. Moreover, this allows us to set a new bound 
for the mass of the electroweak monopole, as we will 
discuss in the following.  

\section{BPS Electroweak Monopoles}

We have shown that, if we replace the bare coupling of 
the standard model by the running coupling, we can 
have a finite energy monopole. This raises an interesting
question. Can we set a BPS bound for the monopole 
with this type of modification? The BPS bound could be 
very useful for the experiments searching for the monopole.
Indeed we can. 

\begin{figure}
\includegraphics[height=4.5cm, width=8cm]{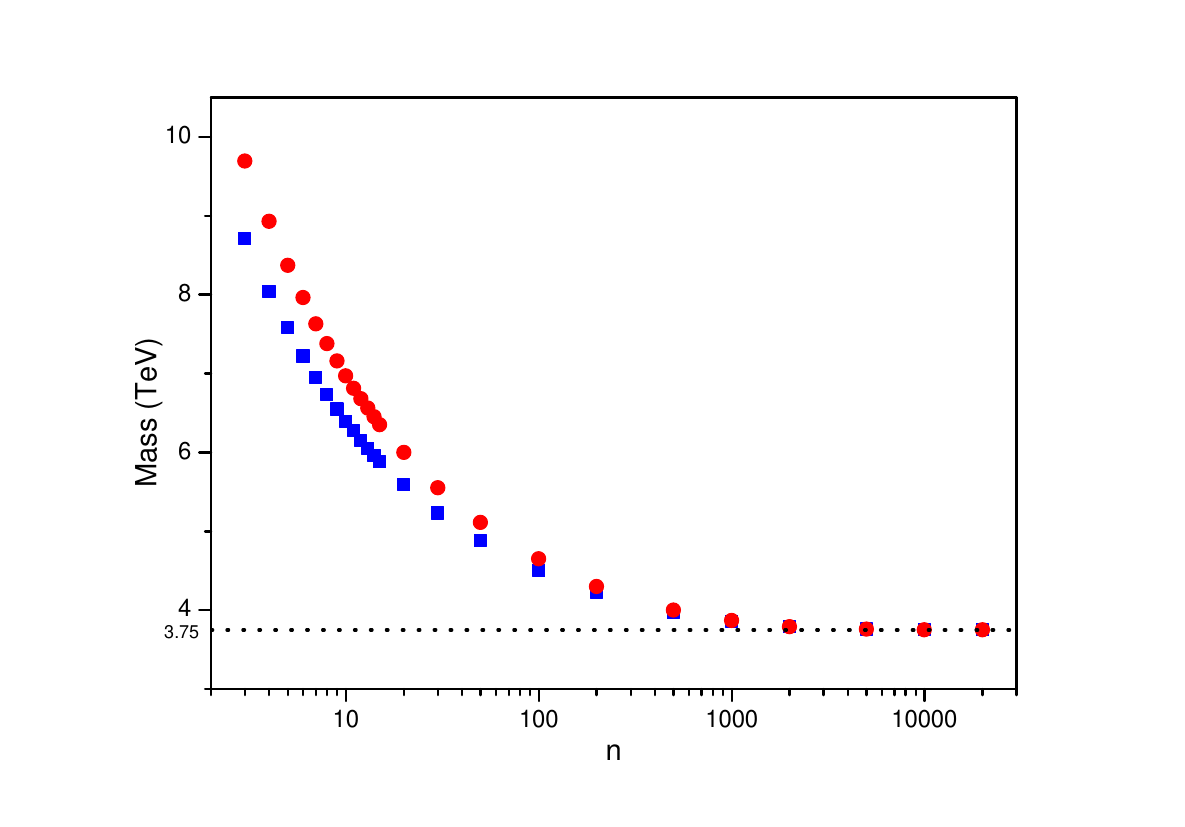}
\caption{\label{mevn} The regularized monopole energy 
in terms of $n$ in log scale. The red dots represent 
the energy given by (\ref{fede2}) with $A=B=0$. 
For comparison we plot the monopole energy given 
by (\ref{fede}) in blue dots. Here the dotted line 
represents the asymptotic energy when $n$ goes 
to infinity.}
\end{figure}

To show this we first consider the following modified 
Lagrangian \cite{bb}
\begin{gather}
{\cal L}= -\frac12 (\pd_\mu \rho)^2 
-\frac{\lambda}{8}\big(\rho^2-\rho_0^2 \big)^2
-\frac{\eps}4 G_\mn^2  \nn\\
-(\frac{\rho_0}{\rho})^2 
\Big\{\frac12 |D'_\mu W_\nu -D'_\nu W_\mu |^2  
-ig F'_\mn W_\mu^* W_\nu   \nn\\
- \frac{g^2}{4}(W_\mu^* W_\nu - W_\nu^* W_\mu)^2 \Big\} 
-\frac{g^2}{4}\rho^2 W_\mu^*W_\mu  \nn\\
-\frac18 \rho^2 (gA_\mu'-g'B_\mu)^2.
\label{bblag}
\end{gather}
Here ${F_\mn'}^2$ term is missing, but this is acceptable
because we are interested in the lower bound of 
monopole energy. Now, with the ansatz (\ref{ans1}) 
the energy of the monopole in the BPS limit is given by
\begin{gather}
E=4\pi \int_0^\infty dr \bigg\{\frac{\eps}{2g'^2 r^2}
+\frac12 (r \dot \rho)^2  
+(\frac{\rho_0}{\rho})^2 \frac{\dot f^2}{g^2} 
+\frac14 \rho^2 f^2 \bigg\}  \nn\\
=4\pi \int_0^\infty dr \bigg\{\frac12 \big(r\dot \rho
-\frac{\sqrt {\eps}}{g' r} \big)^2
+\frac14 (\frac2g \frac{\rho_0}{\rho} \dot f +\rho f)^2  \nn\\
+\frac{\sqrt{\eps}}{g'} \dot \rho  
- \frac{\rho_0}{g}  \dot f f \bigg\}  \nn\\
\geq 4\pi \int_0^\infty dr \Big(\frac{\eps}{g'^2 r^2}
+\frac{\rho^2 f^2}{2} \Big).  
\label{bbme}
\end{gather}
So, when monopole satisfies the Bogomol'nyi equation 
\begin{gather}
\dot \rho-\frac{\sqrt{\eps}}{g' r^2}=0,
~~~~\dot f +\frac{g}{2} \frac{\rho}{\rho_0} \rho f=0,
\label{bbmeq}
\end{gather}
it has the Bogomol'nyi bound. 

\begin{figure}
\includegraphics[height=4.5cm, width=8cm]{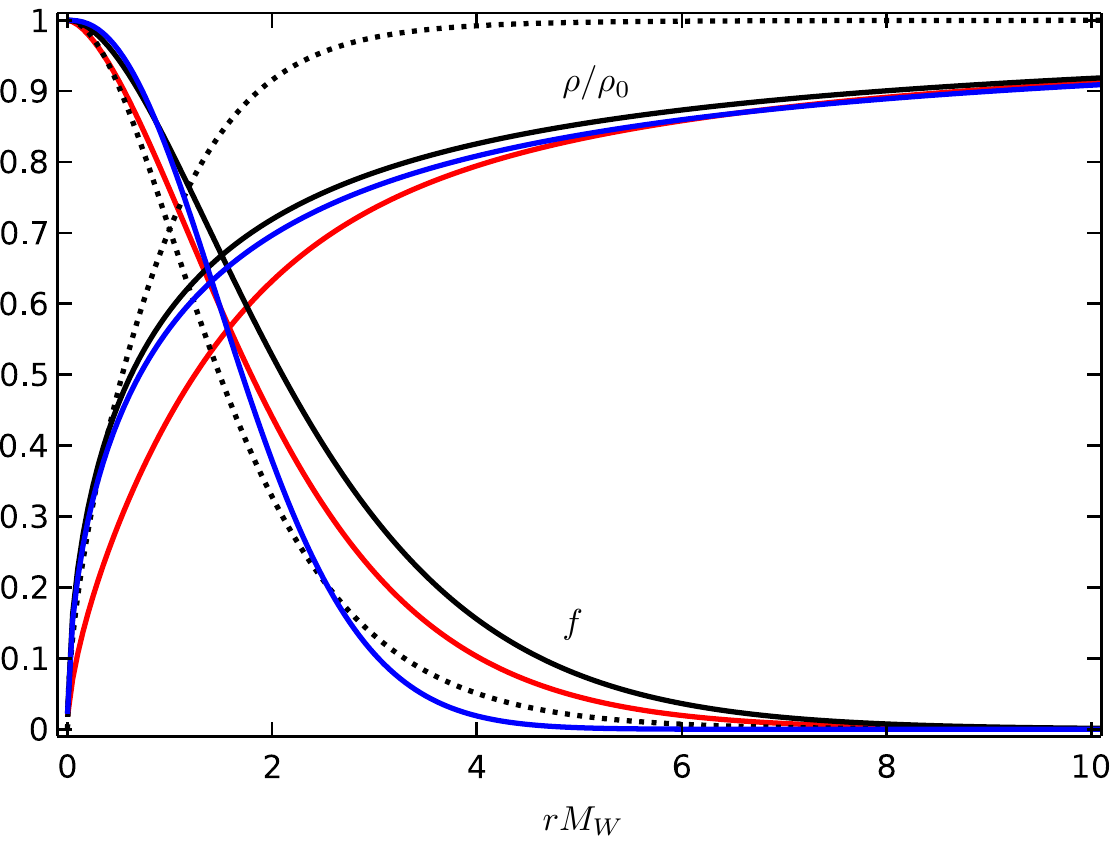}
\caption{\label{bpsm} The BPS electroweak 
monopole solutions. The black curves represent 
the solution given by (\ref{bbm}), the blue curves 
represent the solution given by (\ref{bpsmeq1}), 
and the red curves represent the solution given 
by (\ref{bpsmeq2}). The monopole solution regularized 
by the real electromagnetic permittivity given by 
(\ref{fedeq2}) is shown in dotted curves for comparison.} 
\end{figure}

Integrating the first equation with $\eps=(\rho/\rho_0)^n$ 
and inserting the result to the the second equation, one can 
solve the Bogomol'nyi equation and obtain the following 
solution \cite{bb}
\begin{gather}
\rho(r)=\rho_0 \Big(1
+\frac{n-2}{2 g'\rho_0 r} \Big)^{-\frac{2}{n-2}},   \nn\\
f(r)= f(0)~\exp \Big[-\frac{g (n-2) \rho_0 r}{2(n+2)} 
\Big(\frac{4g' \rho_0 r}{n-2} \Big)^{\frac{4}{n-2} }  \nn\\
\times _2F_1 \Big(1+\frac{4}{n-2}, \frac{4}{n-2}; 
2+\frac{4}{n-2};-\frac{4g' \rho_0 r}{n-2} \Big) \Big],
\label{bbm}
\end{gather} 
where $_2F_1$ is the hypergeometric function. 
The solution with $n=6$ and $f(0)=1$ is shown in 
Fig. \ref{bpsm} in black curves. 

It has been shown that, if one neglects the $\eps$ 
term in (\ref{bbme}) one has the BPS energy bound 
given by \cite{bb}
\begin{gather}
E \geq \frac{2\pi}{g} \rho_0
=\frac{4\pi}{e^2} \sin^2 \theta_{\rm w}~M_{\rm w}    \nn\\
\simeq 2.37~{\rm TeV}.
\label{bbb}
\end{gather}
One might try to obtain a better bound including 
the $\eps$ term. With $\eps=(\rho/\rho_0)^n$  we 
have from 
(\ref{bbmeq})
\begin{gather}
E \geq 4\pi \int_0^\infty \frac{\rho_0}{g} f \dot f dr
+\frac{4\pi}{g'} \int_0^\infty 
\Big(\frac{\rho}{\rho_0} \Big)^{n/2} d \rho   \nn\\
= 4\pi \Big(\frac{1}{2g} 
+\frac{2}{n+2} \frac1{g'} \Big) \rho_0   \nn\\
=\frac{4\pi}{e^2} \sin^2 \theta_{\rm w} 
\Big(1+\frac4{n+2} \cot \theta_{\rm w} \Big) M_{\rm w}.
\end{gather}
But this bound approaches to the above bound as 
$n$ goes to infinity.  

Now, we discuss a more realistic way to set the BPS 
bound for the electroweak monopole. Consider 
the Lagrangian (\ref{lag3}) modified by the real 
electromagnetic permittivity. Compared with 
the Lagrangian (\ref{bblag}), this has a ``minimum" 
modification of the standard model because (\ref{lag3}) 
becomes the Weinberg-Salam Lagrangian (\ref{lag0}) 
with $\beps=1$. This means that, if (\ref{lag3}) has 
the BPS limit, the solution becomes more reliable
than the BPS solution given by (\ref{bblag}).  

To obtain the BPS monopole solution from (\ref{lag3})
notice that, with $A=B=0$ the monopole energy 
(\ref{fede2}) in the BPS limit can be expressed by
\begin{gather}
E =4\pi \int_0^\infty dr \bigg\{\frac{\beps_1}{2e^2 r^2}  
+\frac12 (r\dot \rho)^2   \nn\\
+\frac{1}{2g^2 r^2}(f^2-1)^2
+\frac1{g^2} \dot f^2 +\frac14 \rho^2 f^2 \bigg\}  \nn\\
=4\pi \int_0^\infty dr \bigg\{\frac12 \Big(r\dot \rho
-\frac{\sqrt{\beps_1}}{e r} \Big)^2 
+\frac{\sqrt{\beps_1}}{e} \dot \rho   \nn\\
+\Big(\frac{\dot f}{g}+\frac12 \rho f \Big)^2
-\frac{1}{g} \rho f \dot f
+\frac{1}{2g^2 r^2}(f^2-1)^2 \bigg\}.   
\label{bpsme1}
\end{gather} 
So when we have the Bogomol'nyi equation 
\begin{gather}
\dot \rho -\frac{\sqrt{\beps_1}}{e r^2} =0,
~~~~\dot f +\frac{g}{2} \rho f=0, 
\label{bpsmeq1}
\end{gather}
we have the BPS energy bound 
\begin{gather}
E \geq 4\pi \int_0^\infty dr \bigg\{\frac{\beps_1}{e^2 r^2} 
+ \frac12 \rho^2 f^2    \nn\\
+\frac{1}{2g^2 r^2}(f^2-1)^2  \bigg\}.
\label{bpseb1}
\end{gather} 
This shows that, with (\ref{epbc}) the energy has 
the BPS bound when $f(0)=1$. Clearly this bound 
is better than (\ref{bbme}). 

Integrating the BPS equation with $f(0)=1$ and 
$n=6$, we obtain the monopole solution shown 
in Fig. \ref{bpsm} in blue curves. This has the minimum 
energy given by 6.26 TeV. But this bound has 
a drawback that it depends on the form of $\beps_1$. 
 
To have a new bound which is independent of $\beps_1$, 
notice that when $n$ goes to infinity, $\beps_1$ 
approaches to zero, so that we can neglect it. 
In this limit, we have 
\begin{gather}
\rho=\rho_0,
~~~f= \exp (-\frac{g}{2} \rho_0 r).
\end{gather}
from (\ref{bpsmeq1}). Moreover, (\ref{bpseb1}) 
can be expressed by
\begin{gather}
E > 4\pi \int_0^\infty dr \bigg\{\frac{\rho_0^2}{2} 
\exp (-g \rho_0 r)  \nn\\
+\frac{1}{2g^2 r^2} \big(\exp (-g \rho_0 r)-1 \big)^2 \bigg\}  \nn\\
\simeq 5.89 ~{\rm TeV}.
\label{bpsmb2}
\end{gather} 
This is interesting. But this result may not be so reliable 
for the following reasons. First, with $\rho=\rho_0$ 
the first term in (\ref{bpseb1}) diverges. Second, 
in the limit $n$ goes to infinity, we must have $\rho=0$ 
for any finite value of $r$.  

Interestingly (\ref{lag3}) allows another BPS bound.
To see this notice that the monopole energy 
(\ref{bpsme2}) can also be expressed by
 \begin{gather}
E=4\pi \int_0^\infty dr \bigg\{\frac{\beps_1}{2e^2 r^2}
+\frac12 \Big(r\dot \rho+\frac{1}{g r} (f^2-1) \Big)^2   \nn\\
+\frac{1}{g^2} (\dot f+\frac{g}{2} \rho f)^2
+\frac{1}{g} \dot \rho (f^2-1) 
-\frac{1}{g} \rho f \dot f \bigg\}.  
\label{bpsme2}
\end{gather}
So, the energy has the following BPS bound 
\begin{gather}
E \geq 4\pi \int_0^\infty dr \bigg\{\frac{\beps_1}{2e^2 r^2} 
+\frac12 \rho^2 f^2   \nn\\
+\frac{1}{g^2 r^2} (f^2-1)^2 \bigg\}, 
\label{bpseb3}
\end{gather}
when we have the following BPS equation
\begin{gather}
\dot \rho +\frac{1}{g r^2} (f^2-1)=0,
~~~\dot f+\frac{g}{2} \rho f=0.
\label{bpsmeq2}
\end{gather}  
This has two interesting features. First, this BPS equation 
is independent of $\beps_1$. Second, it looks remarkably 
similar to the Prasard-Sommerfeld equation in 
Georgi-Glashow model. The only difference is the factor 
$1/2$ in the second equation. Integrating (\ref{bpsmeq2}) 
with $n=6$ we obtain the BPS monopole solution 
shown in Fig. \ref{bpsm} in red curves, which has 
the energy 3.96 TeV. 

In the limit $n$ goes to infinity, we can neglect the first 
term in (\ref{bpseb3}) and set a $\beps_1$ independent 
BPS bound 
\begin{gather}
E >  4\pi \int_0^\infty dr \bigg\{\frac12 \rho^2 f^2 
+\frac{1}{g^2 r^2} (f^2-1)^2 \bigg\}   \nn\\
=0.269 \times \frac{4\pi}{e^2} M_{\rm w}  
\simeq 2.98~{\rm TeV}.
\label{bpsmb3}
\end{gather}
This bound provides a new BPS bound for 
the electroweak monopole mass based on (\ref{lag3}), 
which improves the existing bound (\ref{bbb}). 
The result strongly implies that the monopole mass 
may not be smaller than 2.98 TeV. 

This bound should be compared to the bound 3.75 TeV 
shown in Fig. \ref{mevn}. The difference comes (at least 
partly) from the fact that here we have assumed 
$\lambda=0$ and neglected Higgs potential energy. 
Another difference is that the bound given by 
Fig. \ref{mevn} is obtained using the full equation of 
motion, not the BPS equation. In this sense the bound 
3.75 TeV could be viewed more realistic, although this
is obtained numerically with $\beps_1=(\rho/\rho_0)^n$. 

So far we have considered the electroweak monopole 
dressed by both W boson and Higgs field. Now, one 
might wonder if we can have a regularized monopole 
solution without the $W$ boson (i.e., without $f$), with 
only the Higgs field. At first glance this appears to be 
a meaningless question, because the W boson is 
an essential ingredient of the standard model. Moreover, (\ref{cdeq}) tells that, without $f$ Higgs field does not 
couple to the monopole. But this becomes a relevant 
question for the following reasons.

First, when we introduce the quantum correction, the Higgs
field does couple to the monopole. This is clear from
(\ref{fedeq}) and (\ref{fedeq2}). So mathematically this 
becomes an interesting question. More importantly, from
the physical point of view this becomes an important issue 
because this translates to the regularization of Dirac 
monopole. 

\begin{figure}
\includegraphics[height=4.5cm, width=8cm]{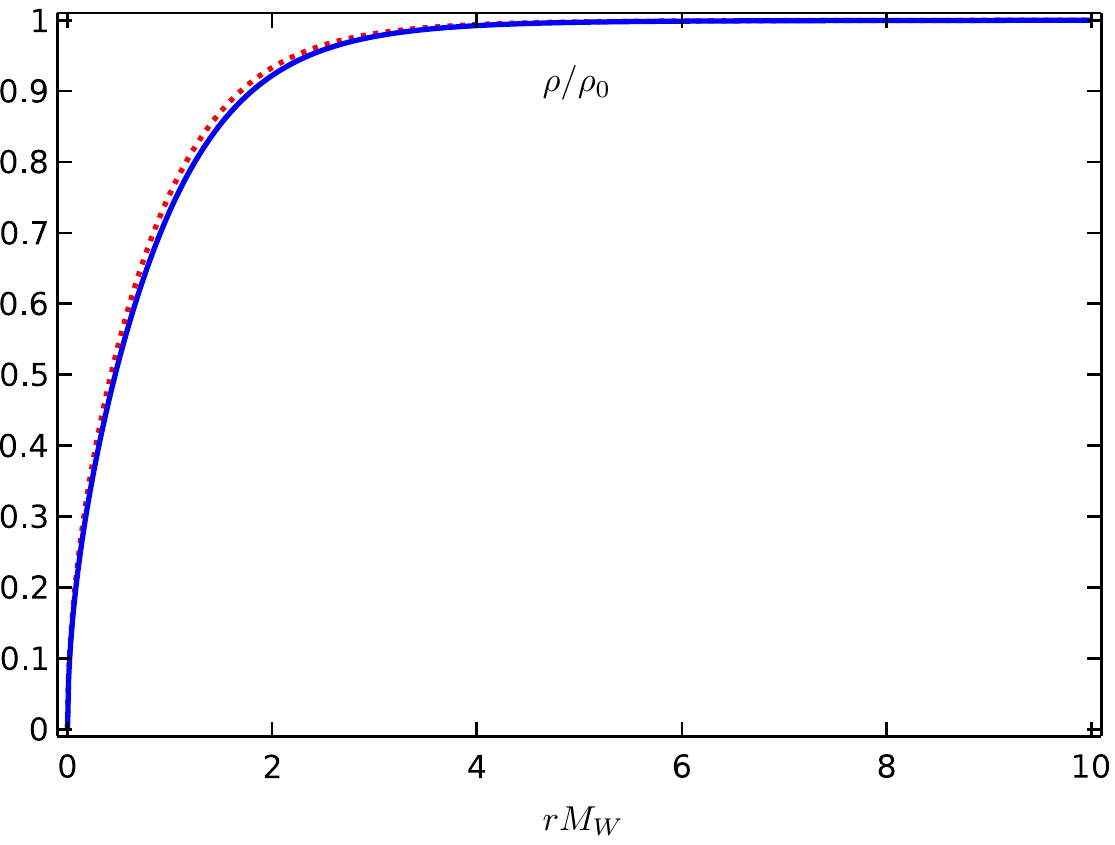}
\caption{\label{femh} The Higgs field dressings of 
the regularized monopole in the absence of the W boson. 
The blue curve corresponds to the real electromagnetic 
permittivity $\beps=(\rho/\rho_0)^6$ and the red curve 
corresponds to the hypercharge permittivity 
$\teps=(\rho/\rho_0)^6$. Exactly the same solution 
also describes the finite energy Dirac monopole 
regularized by the vacuum polarization.}
\end{figure}

It is well known that a major difficulty to find the Dirac 
monopole experimentally has been that there was no 
theoretical estimate of the mass, so that people had 
no idea where to look at the monopole. Now, it must be 
clear that in the absence of the $W$ boson the electroweak 
monopole looks very much like the Dirac monopole. So,
if we could regularize the electroweak monopole in 
the absence of the $W$ boson, we might apply this 
regularization to estimate the mass of the Dirac 
monopole. This makes the  regularization of 
the electroweak monopole in the absence of 
the $W$ boson a very interesting problem.

To find the electroweak monopole solution in the absence 
of the W boson, consider (\ref{fedeq}). In this case 
we can integrate (\ref{fedeq}) and find the the finite 
energy monopole solution which has the Higgs field 
dressing. This is shown in Fig. \ref{femh} in red curve. 

Similarly, we can find the monopole solution regularized 
by the real electric permittivity integrating (\ref{fedeq2}). 
In this case, we have to choose $\beps_0=0$ to make 
the energy (\ref{fede2}) finite. The Higgs field dressing 
of the regularized monopole solution shown in 
Fig. \ref{femh} in blue curve. Remarkably the solution 
is almost identical to the red curve regularized by $\eps$. 
From (\ref{fede1}) we find the energy of this monopole 
for $n=6$ to be around 
\begin{gather}
E \simeq 0.46 \times \frac{4\pi}{e^2}~M_{\rm w}
\simeq 5.09~{\rm TeV}.
\end{gather}
This shows that we can indeed regularize 
the electroweak monopole without the $W$ boson 
with the real electromagnetic permittivity which 
describes the electric charge screening. One may 
wonder if we can generalize the above monopole 
solution to a dyon, integrating (\ref{fedeq2}) with $f=0$. 
But we find that there is no solution which has non-trivial 
$A$ and $B$. 

\section{Regularization of Dirac Monopole}

Since the standard model, in the absence of the W-boson
reduces to the electrodynamics, we may apply the above 
regularization of the electroweak monopole to regularize 
the Dirac monopole. To show this we start from the following electromagnetic U(1) gauge theory coupled to the neutral 
scalar field $\rho$,
\begin{gather}
{\cal L} =-\frac12 (\pd_\mu \rho)^2 
-\frac{\lambda}{8}\big(\rho^2-\rho_0^2 \big)^2
-\frac{1}{4} \beps(\rho)~F_\mn^2. 
\label{elag}
\end{gather}
Here $F_\mn$ is the real electromagnetic field and $\rho$ 
is an emergent scalar field which represents the density 
of electron-positron pairs responsible for the charge 
screening. Obviously this Lagrangian is mathematically 
identical to the Lagrangian (\ref{lag2}) in the absence of 
the $W$ and $Z$ bosons. But from the physical point of 
view we emphasize that this Lagrangian is completely 
independent of the standard model. In particular, here 
$\rho$ is not the Higgs field which makes the photon 
massive, so that $\rho_0$ does not represent the vacuum 
expectation value of the Higgs field in the standard model.  

Now, choose the following ansatz for the Dirac monopole
\begin{gather}
\rho=\rho(r),   \nn\\
A_\mu =A(r) \pd_\mu t 
-\frac{1}{2e} (1-\cos\theta)\pd_\mu \varphi.
\label{ddans}
\end{gather}
With this we have the monopole equation of motion
\begin{gather}
\ddot \rho + \frac{2}{r}\dot \rho
=\frac{\lambda}{2} (\rho^2- \rho_0^2) \rho   
+\frac{\beps'}{2 e^2} 
\Big(\frac{1}{r^4}-\dot A^2 \Big),   \nn\\
\ddot A + \Big(\frac{2}{r} 
+\frac{\beps'}{\beps}\dot \rho \Big) \dot A=0.
\label{ddeq}
\end{gather}
which becomes exactly identical to (\ref{fedeq2}) when 
$f=0$ and $A=B$. This means that mathematically 
(\ref{fedeq2}) in the absence of $W$ and $Z$ bosons 
describes the monopole described by (\ref{ddeq}).

This means that mathematically the monopole solution 
identical to the one shown in Fig. \ref{femh} can describe 
the regularized Dirac monopole dressed by the Higgs 
field in QED. But from the physical point of view this 
monopole is completely different. In particular here 
the monopole energy is expressed by 
\begin{gather}
E=4\pi \int_0^\infty dr \bigg\{\frac{\beps}{2e^2 r^2} 
+\frac12 (r\dot \rho)^2   \nn\\
+\frac{\lambda r^2}{8}\big(\rho^2-\rho_0^2 \big)^2 \bigg\},
\label{dde}
\end{gather}
which, in the limit $\lambda$ goes to zero, is given by
\begin{gather}
E \simeq 0.25 \times \frac{4\pi}{e}~\rho_0,
\label{dde1}
\end{gather} 
where $\rho_0$ now is in principle arbitrary, not related to 
the $W$ boson mass.  

The result in this section is unexpected, because 
it confirms that we can regularize not only the electroweak 
monopole but also the Dirac monopole, replacing the vacuum electromagnetic permittivity with a real electromagnetic 
permittivity. As far as we understand there has been no 
mechanism which makes the energy of Dirac monopole 
finite. Our result shows that the electric permittivity
in real matters could regularize the Dirac monopole.
 
It must be emphasized that the regularization of Dirac 
monopole is independent of the regularization of 
the Cho-Maison monopole in the standard model. 
In particular, $\rho_0$ in (\ref{elag}) is arbitrary (not 
related to the vacuum expectation value of Higgs field 
in the standard model, so that the mass of the Dirac 
monopole given by (\ref{dde1}) is completely independent 
of the mass of the electroweak monopole. Indeed, in 
condensed matters $\rho_0$ could easily be of the order 
of meV, the vacuum expectation value of the Higgs field in 
Landau-Ginzburg theory. This implies that a regularized 
Dirac monopole of mass around hundred meV could 
exist in real condensed matters. 

\section{Conclusion}

The fact that the energy of the Cho-Maison monopole 
is infinite has made some people to doubt the existence 
of the monopole in the standard model. But we emphasize 
that the existence of the monopole is determined by 
the topology of the standard model, not the classical
energy of the monopole \cite{plb97,yang}. Moreover, 
since the standard model has the monopole topology, 
the electroweak monopole must exist.  

For this reason the search for the electroweak monopole
has been taken seriously \cite{medal,atlas,icecu}. A crucial 
piece of information needed to make the search successful 
is the monopole mass. Generally speaking, the mass is 
expected to be around 4 to 10 TeV \cite{epjc15,ellis,bb}, 
but we need a more precise prediction. 

In this paper we showed that indeed the electromagnetic 
permittivity can regularize the electroweak monopole. 
This has deep implications. First of all, this shows that 
there is a more natural way to regularize the monopole,
which makes the electroweak monopole more realistic. 
Perhaps more importantly, this sets a new limit on  
the monopole mass. Indeed, our result strongly implies 
that the mass of the monopole may not be smaller than
3 TeV (more realistically 3.75 TeV). This could be 
an important information for experiments to detect 
the electroweak monopole, in particular the MoEDAL 
and ATLAS experiments at CERN, because the present 
14 TeV LHC could produce the monopole only when 
the monopole mass is less than 7 TeV \cite{medal,atlas}.  

Furthermore, our result shows that the charge screening 
could also regularize the Dirac monopole. This could 
revitalize the search for the Dirac monopole. One reason 
why the experimental search for the Dirac monopole 
has not been so successful so far is that it has been 
a blind search in the dark room, without any theoretical 
hint on the mass \cite{cab}. But a more serious reason 
is that the Dirac monopole may not exist in nature as 
a fundamental particle, because the electroweak unification transforms it to the electroweak monopole. 

But there is no reason why this monopole could not exist 
as a quasi-stable particle. To understand this point, 
consider the well known Abrikosov vortex in ordinary superconductors \cite{abri}. Although this vortex does 
not exist as a fundamental object, we can create it in 
condensed matters applying the magnetic field from 
the outside. Exactly the same way, we could create 
the Dirac monopole in condensed matters by brute force, 
applying a spherically symmetric magnetic field. Certainly
this would not be easy, but there is no reason why this 
could not be done.

This strongly implies that for the Dirac monopole what is
important is creating it in condensed matters, rather than 
searching for the existing one in nature. Our analysis 
shows that in this case the mass of the Dirac monopole 
could be of the order of 100 meVs, since $\rho_0$ in 
condensed matters is expected to be of the order of 
meV. This is remarkable. 

{\bf ACKNOWLEDGEMENT}

~~~The work is supported in part by the National Research 
Foundation of Korea funded by the Ministry of Education 
(Grants 2015-R1D1A1A0-1057578, 2018-R1D1A1B0-7045163), 
National Natural Science Foundation of China (Grant 11975320
and Grant 11805242), and by Center for Quantum Spacetime, 
Sogang University, Korea.

\end{document}